\documentclass[aps,prl,reprint,groupedaddress]{revtex4-1}

\newcommand{\ep}{\textit{ep}}
\newcommand{\etal}{{\it et~al}.}
\usepackage{graphicx}
\usepackage{dcolumn}
\usepackage{bm}
\usepackage{color}
\usepackage{array}
\usepackage{footmisc}

\begin{document}


\title{Insensitivity of sub-Kelvin electron-phonon coupling to substrate properties}


\author{Jason M. Underwood}
\author{Peter J. Lowell}
\author{Galen C. O'Neil}
\author{Joel N. Ullom}
\affiliation{National Institute of Standards and Technology, Boulder, CO USA}



\begin{abstract}
We have examined the role of the substrate on electron-phonon coupling in normal metal films of Mn-doped Al at temperatures below 1~K. Normal metal-insulator-superconductor junctions were used to measure the electron temperature in the films as a function of Joule heating power and phonon temperature. Theory suggests that the distribution of phonons available for interaction with electrons in metal films may depend on the acoustic properties of the substrate, namely, that the electron-phonon coupling constant $\Sigma$ would be larger on the substrate with smaller sound speed. In contrast, our results indicate that within experimental error (typically $\pm 10$~\%), $\Sigma$ is unchanged among the two acoustically distinct substrates used in our investigation. 
\end{abstract}


\maketitle


At temperatures below 1~K, the interaction between electrons and phonons in metals weakens to the point that the two systems can be out of thermal equilibrium with one another. Such weakened coupling may be undesirable e.g., in dc superconducting quantum interference devices (SQUIDs) where the noise may be limited by hot-electron effects in the normal-metal shunts~\cite{Wellstood:1988}. This effect is also the origin of the problem of low-temperature thermometry wherein the electrons of the thermometer are hotter than the surrounding material. On the other hand, such decoupling may be desirable to increase the performance of electron-tunneling microrefrigerators based on normal metal-insulator-superconductor (NIS) junctions. Electron-phonon decoupling can also be exploited to make cryogenic bolometers where dynamic range and sensitivity are strong functions of the decoupling level~\cite{Wei:2008}.

Electron-phonon (\ep) coupling occurs because passing phonons distort the local lattice structure, and conduction electrons respond to the resulting band distortions. A useful approximation for this interaction is the scalar deformation potential~\cite{Kittel:1996}, which relates the change in Fermi level $\delta \varepsilon_F$ to the local volume change or dilation, $\Delta V/V$, of a unit cell,
	\begin{equation}
	\delta \varepsilon_F = \varepsilon_F- \varepsilon_F^0 = -\frac{2}{3} \varepsilon_F \frac{\Delta V}{V}.
	\end{equation}
For a longitudinal phonon, the dilation magnitude $|\Delta V|/V = \sqrt{\hbar q/2\rho\Omega c}$, where $\hbar$ is the reduced Planck's constant, $q$ the phonon wavevector, $\rho$ the mass density, $\Omega$ the crystal volume, and $c$ the sound speed. A more refined theory of \ep\ coupling, accounting for interaction between electrons and transverse phonons, was developed by Reizer~\cite{Reizer:1989}. Later, Sergeev and Mitin~\cite{Sergeev:2000} extended Reizer's work by accounting for quantum interference that occurs when static impurities are present in the host metal. Regardless of the framework employed, the two important quantities derived in these theories are the electron relaxation rate $\tau_{ep}^{-1}$, and the heat flow, or thermopower, between the electron and phonon subsystems $P_{ep}$. 

Calculation of $P_{ep}$ involves integration over the electron states $\mathbf{k}$ and phonon states $\mathbf{q}$ of an electron-phonon collision term $\mathcal{M}_{ep}$, 
	\begin{equation}\label{epIntegral1}
	P_{ep} \sim \int d\mathbf{k} \int d\mathbf{q} \; \mathcal{M}_{ep}(\mathbf{k},\mathbf{q},T_e,T_p), 
	\end{equation}
where $T_e$ ($T_p$) is the electron (phonon) temperature. Converting to integration over energies $\omega(\mathbf{q})$ and $\varepsilon(\mathbf{k})$, we have
	\begin{equation}\label{epIntegral2}
	P_{ep} \sim \int d\varepsilon \int d\omega \, \mathcal{D(\omega)} \mathcal{M}_{ep}(\varepsilon,\omega,T_e,T_p),
	\end{equation}
where $\mathcal{D(\omega)}$ is the phonon density of states. Since $\mathcal{M}_{ep}$ contains factors of the form $f_e(1-f_e)$, where $f_e(\varepsilon,T_e)$ is the Fermi-Dirac distribution function, only the electron density of states at the Fermi level $\nu(\varepsilon_F)$ is relevant, and this can be brought out of the $\varepsilon$ integral. Because of the factors $\nu(\varepsilon_F)$ and $\mathcal{D(\omega)}$, as well as deformation potential-like terms in $\mathcal{M}_{ep}$, one finds that $P_{ep}$ is principally dependent on the Fermi velocity $\upsilon_F$ and the sound speed $c$ of the metal. 

For bulk metals the \ep\ theory outlined above is straightforward enough. However, many experimental situations that rely on \ep\ coupling involve the use of thin metal films on dielectric substrates, and the correct form to use for $\mathcal{D(\omega)}$ in Eq.~(\ref{epIntegral2}) is not always clear. At low temperatures it is possible to be in a regime such that the thickness $d$ of a film is much less than the dominant thermal phonon wavelength $\lambda_\mathrm{ph} \approx hc/4k_{B}T$, where $k_B$ is Boltzmann's constant, and the 4 in the denominator is based on an empirical conversion factor found in Ref.~\onlinecite{Klitsner:1987}. For a free-standing film, with $\lambda_\mathrm{ph} \gg d$, the phonon distribution becomes two-dimensional (2D) and the phonon density of states changes to reflect this new dimensionality, as well as the presence of flexural Rayleigh-Lamb modes~\cite{Karvonen:2007}. On the other hand, if the film is strongly adhered to a rigid substrate, then the proper system to consider is that of the composite film/substrate, which should exhibit a three-dimensional (3D) phonon spectrum. Further, for $\lambda_\mathrm{ph} \gg d$ it may be better justified to characterize that 3D phonon distribution with substrate parameters (e.g., $\rho_\mathrm{sub}$ and $c_\mathrm{sub}$), rather than those of the metal film. We report in this Letter the dependence of \ep\ coupling on substrate, by comparing the measured \ep\ thermopower in normal metal thin films of Al doped with 4200~ppm Mn~\cite{ONeil:2010} on two substrates (Si and SiO$_2$) with very different sound speeds (cf. Table~\ref{SoundSpeedTable}). 

\begin{table}[htbp]
\caption{\label{SoundSpeedTable}Longitudinal and transverse phonon speeds in [m/s] for materials used in this work. Also shown is the expected trend for $\Sigma$, based on electronic parameters of the Al:Mn film and sound speeds for each substrate.}
\begin{ruledtabular}
\begin{tabular}{>{\hspace{0.1in}} c <{\hspace{0.1in}} | c <{\hspace{0.1in}} | c <{\hspace{0.1in}} | c <{\hspace{0.1in}} }
substrate  & 	longitudinal	& 	transverse   & 	expected $\Sigma$ \\
\hline
Al 	& 	 6380 	&   3070  	&	$ \Sigma_\mathrm{Al}$ 	\\
SiO$_2$ 	& 	  5840 & 	 3700  & $\approx \Sigma_\mathrm{Al}$ \\
Si	& 9420 & 	5860  &  $\ll \Sigma_\mathrm{Al}$ \\ 
\end{tabular}
\end{ruledtabular}
\end{table}

In order to quantify our results, we assumed the following form for the \ep\ thermopower:  
	\begin{equation}\label{epPower}
	P_{ep} = \Sigma \Omega \left(T_e^n - T_p^n \right) \\,
	\end{equation}
where $\Omega$ is the normal-metal volume, $n$ = 4 to 6, and $\Sigma$ is the material-dependent \ep\ coupling constant of order 1~nW/$\mu$m$^3$K$^n$. This expression for $P_{ep}$ follows from completing the integration in Eq.~(\ref{epIntegral2}). Depending on the particular regime (e.g., amount of disorder, temperature) theory predicts different expressions for $\Sigma$ and different values for $n$~\cite{Reizer:1989,Sergeev:2000}. The different regimes can be parameterized by the product of the dominant thermal phonon wavevector $q_T = 4k_{B}T/\hbar c$ and the elastic electron mean free path $\ell$, as depicted in Table~\ref{epRegimes}. 

\begin{table}[htbp]
\caption{\label{epRegimes}Electron-phonon coupling regimes in bulk normal metals.}
\begin{ruledtabular}
\begin{tabular}{>{\hspace{0.1in}\bfseries} l <{\hspace{0.1in}} | c <{\hspace{0.3in}} | c <{\hspace{0.25in}} }
$q_T \ell \rightarrow \infty$  & 	pure limit	& 	$n$ = 5 \\
$q_T \ell > 1$	& clean limit & $4 < n < 5$  \\ 
$q_T \ell < 1$ 	& 	  impure/dirty limit & $4 < n < 6$  \\
\end{tabular}
\end{ruledtabular}
\end{table}

The variation in $n$ for $q_T \ell \sim 1$ is due to quantum interference between ``pure'' electron-phonon and electron-boundary/impurity scattering. In the clean limit, $n \rightarrow 4$ if \ep\ interaction is dominated by transverse phonons and scatterers have a static character --- e.g., massive impurities or rigid boundaries. The same behavior is found in the dirty limit as the static component of impurities increases. In other words, an increase in the fraction $\delta$ of static scatterers enhances \ep\ coupling ($n \rightarrow 4$), while an increase in the overall concentration of scatterers weakens \ep\ coupling ($n \rightarrow$ 5 or 6). The theory developed in Ref.~\onlinecite{Sergeev:2000} for the dirty limit yields an expression for $P_{ep}$ that is actually the sum of $n=4$ and $n=6$ terms,
\begin{equation}\label{epPower2}
	P_{ep} = \Sigma_4 \Omega \left(T_e^4 - T_p^4 \right) + \Sigma_6 \Omega \left(T_e^6 - T_p^6 \right). 
\end{equation}
As a result, when we fit the data to Eq.~(\ref{epPower}) we will obtain a single value for $n$ that is between 4 and 6, depending on the relative contribution of the two terms in Eq.~(\ref{epPower2}). A similar expression exists for the clean limit, but with $n = 5$ instead of 6 and a different formula for $\Sigma_4$. The reason for fitting the data to Eq.~(\ref{epPower}), and not Eq.~(\ref{epPower2}) is that the theoretical expressions for $\Sigma_n$ apply only to their respective limits, and as we discuss later, our data cover the crossover between the impure and clean regimes. The full expressions for $\Sigma_n$ are complicated, but they all depend on the sound speed $c$ as 
\begin{equation}\label{SigmaVsSpeed}
	\Sigma_n \sim \frac{1}{c^{n-1}}. 
\end{equation}
Therefore, if the substrate affects the phonon density of states $\mathcal{D}(\omega)$ in the metal film, then the measured $\Sigma_n$ should differ by a factor in the range of 4--10 on the two substrates we have chosen (cf. Table~\ref{SoundSpeedTable} and Eqs.~(\ref{epPower2}) and (\ref{SigmaVsSpeed})). 

In order to measure $\Sigma$, we fabricated thin-film resistors (thickness $d = 35$~nm) with integrated NIS (normal metal-insulator-superconductor) thermometers (see inset of Fig.~\ref{IV_100mK}), using standard photolithographic processing~\cite{[See Supplemental Material at ][ for additional details related to device fabrication and experimental setup.]supplement}. For our Al:Mn films and the temperature range covered, the dominant transverse phonon wavelength $\lambda_\mathrm{ph} =$ 60--370~nm, which is always greater than $d$. NIS junctions provide reliable thermometry as long as the product of the normal-state tunnel resistance $R_T$ and junction area is greater than about 10~k$\Omega$-$\mu$m$^2$. For these high-resistance junctions, normal-metal self-cooling is negligible, and the local electron temperature near the thermometer junction is identical to that in the rest of the normal-metal film. Heater current was injected into the resistor via superconducting electrodes. The resulting NS contacts provide very good electrical conductance, but very poor thermal conductance, since they are always biased within the superconducting gap $\Delta_\mathrm{Al}$. 

\begin{figure}[htb]
\includegraphics[width=3.38in]{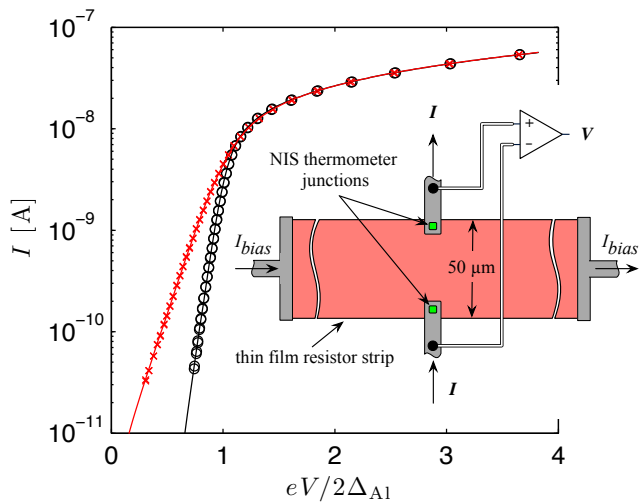}
\caption{\label{IV_100mK}NIS thermometer $IV$ data (symbols) and corresponding fits (lines) for $T_\mathrm{bath} \approx 0.1$~K. Black circles represent the unheated ($P=0$) data, while red crosses indicate the data for an applied heater power $P = 740$~pW. The temperatures of the two fits are $T=118$~mK ($P=0$) and $T=275$~mK ($P=740$~pW). Inset: Schematic of NIS thermometer/heater device and measurement arrangement. Red resistor strip is Al:Mn. Grey current injection electrodes are superconducting Al. Resistor length is 500~$\mu$m.}
\end{figure}

Since we are ultimately interested in determining the parameters $\Sigma$ and $n$, and since the temperature dependence of \ep\ coupling is roughly $T^5$, accurate determination of temperature is critical. The current-voltage ($IV$) curves of our NIS thermometers were found to fit rather well (cf. Fig~\ref{IV_100mK}) to the standard isothermal theory, based on the BCS density of states \cite{Tinkham:2004}, 
	\begin{equation}\label{BCStheory}
	I(V) = \frac{1}{eR_T}\int dE \frac{E}{\sqrt{E^2-\Delta^2}}\left[f(E)-f(E+eV) \right],
\end{equation}
where $e$ is the elementary charge and $f(E)$ is the Fermi-Dirac distribution. Using the measured energy gap of Al ($\Delta_\mathrm{Al} \approx 185$~$\mu$eV) and the measured tunneling resistance $R_T \approx 6$~k$\Omega$ (per junction) as inputs, and electron temperature $T_e$ as a fitting parameter, we fit $IV$ data to Eq.~(\ref{BCStheory}) by use of a nonlinear least-squares algorithm. A logarithmic weighting was applied to the current ($I$) data, since the region of the $IV$ curve below the gap has the greatest temperature response. This procedure allowed us to avoid some of the  calibration difficulties associated with resistance thermometry. 

The $IV$ curve fit at zero power determines the bath temperature $T_\mathrm{bath}$, which is nearly equal to the cryostat temperature, as expected when stray loads are negligible. By sweeping over all bath temperatures and applied powers $P$, and fitting the respective $IV$ curves, we obtained a data set $T_{e}(P,T_\mathrm{bath})$. The phonon temperature $T_p$ in the film was calculated from a model for power flow out of the phonon system that includes the predicted Kapitza resistance between the film and substrate~\cite{Swartz:1989} and the predicted thermal conductivity of the substrate itself. Thermal resistance between the substrate and copper mounting box was not included in this model, but it was estimated to be much smaller than that between film and substrate. Ultimately, the calculated differences between $T_p$ and $T_\mathrm{bath}$ are very small, so our final results are insensitive to the details of this model. The final data set $T_{e}(P,T_{p})$, along with our measurement of the heater strip volume, were used to determine $\Sigma$ and $n$ by fitting to Eq.~(\ref{epPower}). 

\begin{figure}[htb]
\includegraphics{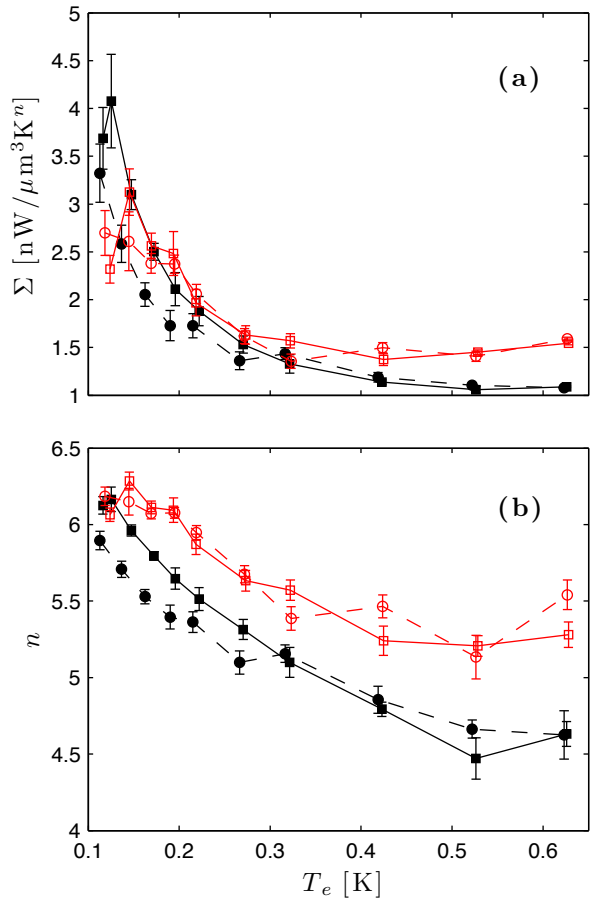}
\caption{\label{SigmaN}(a) Measured $\Sigma$ and (b) $n$ vs.\ electron temperature $T_e$ for Al:Mn resistors on Si (red, open symbols) and SiO$_2$ (black, filled symbols) substrates. The dashed and solid lines for each color represent different samples from the same wafer. Error bars indicate $\pm$ one standard deviation in the uncertainty of the fit.}
\end{figure}

The primary result of this Letter is presented in Fig.~\ref{SigmaN}(a). Aside from a slight deviation among the two substrates at higher temperatures, it is readily apparent that changing substrates has a negligible effect on the \ep\ coupling constant. The overall magnitude of $\Sigma$ is consistent with previous measurements on Al:Mn thin films~\cite{Taskinen:2006}. The variation in $\Sigma$ with temperature likely indicates a crossover from the impure limit at low $T$ to the clean limit at high $T$. For example, assuming $q_T$ is a function only of the transverse sound speed in Al, we find $0.2 < q_T \ell < 1.4$ over the temperature range 0.1~K to 0.6~K for films on Si. Similarly, $0.4 < q_T \ell < 2.6$ for films on SiO$_2$. For the temperature range explored here, transverse phonons dominate the interaction with electrons. The characteristic temperature $T_1^*$ for transverse-longitudinal crossover can be estimated by comparing the electron relaxation rates of each polarization~\cite{Sergeev:2000} in the regime $q_T \ell \sim 1$, 
	\begin{equation}
	T_1^* = \frac{6\pi}{7\zeta(3)}\frac{\hbar c_l}{k_B \ell} \left(\frac{c_l}{c_t}\right)^3, 
	\end{equation}
where $\zeta(3) \approx 1.202$ and $c_l$ ($c_t$) is the longitudinal (transverse) sound speed. For Al:Mn films on Si, $T_1^* \approx 70$~K. Because we are primarily concerned with the sub-kelvin regime, we can safely assume that transverse phonons are much more important for \ep\ coupling. 

The clean-impure crossover is made more clear when considering the variation of the \ep\ exponent $n$ with $T_e$, as shown in Fig.~\ref{SigmaN}(b). Here, it would seem that the substrate does exhibit an influence, but we hypothesize that the difference in $n$ between the two substrates can be attributed to a difference in the fraction of static scatterers $\delta \sim \ell/ \mathcal{L}$~\cite{Sergeev:2000}, where $\mathcal{L}$ is the mean free path due to static scattering. If we assume that scattering at the upper and lower film surfaces is predominantly static~\footnote{Static scattering at film boundaries may result from lattice distortions due to strain or reconstruction at the film surface and thermal expansion mismatch at the film-substrate interface}, the film thickness $d$ sets the scale for $\mathcal{L}$, and $\delta \sim \ell/d$. For our films on SiO$_2$ substrates, $\ell_\mathrm{SiO_2} \approx 25$~nm, which is nearly twice that of our films on Si ($\ell_\mathrm{Si} \approx 14$~nm)~\cite{supplement}. Since $d$ is the same for both films, $\delta_\mathrm{SiO_2} \approx 2 \delta_\mathrm{Si}$, and we expect that Al:Mn films on SiO$_2$ should indeed exhibit a smaller exponent than the films on Si, at a given temperature. 

The insensitivity of the \ep\ coupling to the substrates studied here presents a puzzling problem. If the phonons of the substrate-film system are pictured as a gas that is able to move freely between the film and substrate, then phonons from the substrate will dominate the instantaneous phonon population in the film, since each body contributes a number of phonons to the gas that is proportional to its volume~\cite{Kittel:1996}. In this picture, substrate properties are expected to strongly influence \ep\ coupling. 

Since $\Sigma$ may increase or decrease with $\ell$~\cite{Sergeev:2000}, one may argue that the dependence of $\Sigma$ on $c$ (i.e., the substrate) is hidden by virtue of the films exhibiting different $\ell$s on the two substrates. However, for $T<0.3$~K, the $n=6$ term in Eq.~(\ref{epPower2}) is dominant for films on either substrate. For this case, $\Sigma_6 \sim \ell/c^5$, and we estimate $\Sigma_{6,\mathrm{SiO_2}}/\Sigma_{6,\mathrm{Si}} \approx 18$. In other words, at low temperatures the difference in $\ell$ should only \emph{enhance} the difference in $\Sigma$ due to $c$. Since the observed $\Sigma$s for the two substrates are practically equal for all temperatures considered here, it is reasonable to assume that the different mean free paths do not impact our overall conclusion that \ep\ coupling in metal films is not strongly dependent on substrate.

Decoupling of the phonon systems in the film and substrate could reduce the dependence of the \ep\ interaction on the substrate. For example, phonon trapping due to acoustic mismatch could lead to modification of the phonon gas dimensionality in the film, despite the supported geometry used in this work. If the sound speed in the metal film is less than that in the substrate, phonons in the film experience total internal reflection when their angle of incidence at the film-substrate boundary is greater than a critical angle $\theta_c = \sin^{-1}(c_\mathrm{film}/c_\mathrm{sub})$ from the film-substrate interface normal~\cite{Kaplan:1979}. This effect is the acoustic analogue to Snell's law in optics. Only phonons that approach the substrate interface from within the cone are able to travel across the interface. Phonons traveling on paths outside the critical cone are trapped within the film and resemble a 2D phonon gas~\cite{Nabity:1991}. For Al:Mn on Si (SiO$_2$), the angle of the critical cone for the dominant transverse phonons is $32^\circ$ (56$^\circ$). In other words, in films with a narrower critical cone (e.g., Al:Mn on Si), phonon trapping is enhanced and the film phonon spectrum may become 2D. However, for real interfaces other phonon scattering mechanisms exist, which tend to make the phonon gas 3D~\cite{Nabity:1991}. 

Although there are many factors (e.g., temperature, disorder) that determine the observed \ep\ exponent $n$, we can still use our data for $n$ to help elucidate whether phonon dimensionality plays a role in our null result.  First, the theory of Sergeev \etal~\cite{Sergeev:2000} --- appropriate for bulk samples exhibiting a 3D phonon spectrum --- predicts that our Al:Mn samples should exhibit an $n$ that varies from 6 to 4, as temperature is increased. This prediction is supported by our \ep\ coupling measurements and those on similar Mn-doped Al films~\cite{Taskinen:2006}. Second, it was indicated both theoretically and experimentally~\cite{*[{See }] [{ and references therein.}] Karvonen:2007} that \ep\ coupling in the presence of a 2D phonon gas is associated with a value of $n \leq 4.5$. Finally, it may be argued that $\lambda_\mathrm{ph}$ is close enough to $d$ that a crossover from 3D to 2D phonons occurs as temperature is lowered. However, the observed trend for Al:Mn films on both substrates is for $n$ to increase as $T$ is decreased. If the film phonons become more 2D as $T$ is lowered, we would expect $n$ to decrease or remain fairly constant. These points strongly suggest that there is little meaningful difference in phonon dimensionality between Al:Mn films on the two substrates and that phonons in the films are closer to 3D than 2D.

It is interesting that we obtain qualitative agreement with the theory of Sergeev \etal~\cite{Sergeev:2000} by assuming the films behave as bulk samples ($\Sigma_6 = $ 4--14~nW/$\mu$m$^3$K$^6$). However, since $\lambda_\mathrm{ph}$ is always greater than the film thickness $d$, it is not possible to construct phonon modes in the direction perpendicular to the film, and therefore yield a 3D phonon gas, without also considering the substrate as part of the overall problem. Our observation that the phonons are 3D, and yet \ep\ coupling is insensitive to substrate, cannot be accounted for within this theoretical framework. 

In summary, we have demonstrated that \ep\ coupling in thin normal-metal films is largely independent of the acoustic properties of the supporting substrate. The results presented have important implications for efforts to engineer low-temperature micro- or nanostructures, which depend on \ep\ coupling for operation. Furthermore, the ubiquity of film-substrate geometries points to the need for models of \ep\ coupling that consider such geometries explicitly. 

\begin{acknowledgments}
This work was supported by a National Research Council Postdoctoral Fellowship and the NASA APRA program. US government contribution; not subject to copyright in the United States.
\end{acknowledgments}


\begin{thebibliography}{15}%
\makeatletter
\providecommand \@ifxundefined [1]{%
 \@ifx{#1\undefined}
}%
\providecommand \@ifnum [1]{%
 \ifnum #1\expandafter \@firstoftwo
 \else \expandafter \@secondoftwo
 \fi
}%
\providecommand \@ifx [1]{%
 \ifx #1\expandafter \@firstoftwo
 \else \expandafter \@secondoftwo
 \fi
}%
\providecommand \natexlab [1]{#1}%
\providecommand \enquote  [1]{``#1''}%
\providecommand \bibnamefont  [1]{#1}%
\providecommand \bibfnamefont [1]{#1}%
\providecommand \citenamefont [1]{#1}%
\providecommand \href@noop [0]{\@secondoftwo}%
\providecommand \href [0]{\begingroup \@sanitize@url \@href}%
\providecommand \@href[1]{\@@startlink{#1}\@@href}%
\providecommand \@@href[1]{\endgroup#1\@@endlink}%
\providecommand \@sanitize@url [0]{\catcode `\\12\catcode `\$12\catcode
  `\&12\catcode `\#12\catcode `\^12\catcode `\_12\catcode `\%12\relax}%
\providecommand \@@startlink[1]{}%
\providecommand \@@endlink[0]{}%
\providecommand \url  [0]{\begingroup\@sanitize@url \@url }%
\providecommand \@url [1]{\endgroup\@href {#1}{\urlprefix }}%
\providecommand \urlprefix  [0]{URL }%
\providecommand \Eprint [0]{\href }%
\providecommand \doibase [0]{http://dx.doi.org/}%
\providecommand \selectlanguage [0]{\@gobble}%
\providecommand \bibinfo  [0]{\@secondoftwo}%
\providecommand \bibfield  [0]{\@secondoftwo}%
\providecommand \translation [1]{[#1]}%
\providecommand \BibitemOpen [0]{}%
\providecommand \bibitemStop [0]{}%
\providecommand \bibitemNoStop [0]{.\EOS\space}%
\providecommand \EOS [0]{\spacefactor3000\relax}%
\providecommand \BibitemShut  [1]{\csname bibitem#1\endcsname}%
\let\auto@bib@innerbib\@empty
\bibitem [{\citenamefont {Wellstood}(1988)}]{Wellstood:1988}%
  \BibitemOpen
  \bibfield  {author} {\bibinfo {author} {\bibfnamefont {F.~C.}\ \bibnamefont
  {Wellstood}},\ }\emph {\bibinfo {title} {Excess noise in the dc SQUID; 4.2K
  to 20mK}},\ \href@noop {} {Ph.D. thesis},\ \bibinfo  {school} {University of
  California, Berkeley} (\bibinfo {year} {1988})\BibitemShut {NoStop}%
\bibitem [{\citenamefont {Wei}\ \emph {et~al.}(2008)\citenamefont {Wei},
  \citenamefont {Olaya}, \citenamefont {Karasik}, \citenamefont {Pereversev},
  \citenamefont {Sergeev},\ and\ \citenamefont {Gershenson}}]{Wei:2008}%
  \BibitemOpen
  \bibfield  {author} {\bibinfo {author} {\bibfnamefont {J.}~\bibnamefont
  {Wei}}, \bibinfo {author} {\bibfnamefont {D.}~\bibnamefont {Olaya}}, \bibinfo
  {author} {\bibfnamefont {B.~S.}\ \bibnamefont {Karasik}}, \bibinfo {author}
  {\bibfnamefont {S.~V.}\ \bibnamefont {Pereversev}}, \bibinfo {author}
  {\bibfnamefont {A.~V.}\ \bibnamefont {Sergeev}}, \ and\ \bibinfo {author}
  {\bibfnamefont {M.~E.}\ \bibnamefont {Gershenson}},\ }\href@noop {}
  {\bibfield  {journal} {\bibinfo  {journal} {Nature Nanotech.}\ }\textbf
  {\bibinfo {volume} {3}},\ \bibinfo {pages} {496} (\bibinfo {year}
  {2008})}\BibitemShut {NoStop}%
\bibitem [{\citenamefont {Kittel}(1996)}]{Kittel:1996}%
  \BibitemOpen
  \bibfield  {author} {\bibinfo {author} {\bibfnamefont {C.}~\bibnamefont
  {Kittel}},\ }\href@noop {} {\emph {\bibinfo {title} {Introduction to Solid
  State Physics}}},\ \bibinfo {edition} {7th}\ ed.\ (\bibinfo  {publisher}
  {John Wiley \& Sons},\ \bibinfo {year} {1996})\BibitemShut {NoStop}%
\bibitem [{\citenamefont {Reizer}(1989)}]{Reizer:1989}%
  \BibitemOpen
  \bibfield  {author} {\bibinfo {author} {\bibfnamefont {M.}~\bibnamefont
  {Reizer}},\ }\href@noop {} {\bibfield  {journal} {\bibinfo  {journal} {Phys.
  Rev. B}\ }\textbf {\bibinfo {volume} {40}},\ \bibinfo {pages} {5411}
  (\bibinfo {year} {1989})}\BibitemShut {NoStop}%
\bibitem [{\citenamefont {Sergeev}\ and\ \citenamefont
  {Mitin}(2000)}]{Sergeev:2000}%
  \BibitemOpen
  \bibfield  {author} {\bibinfo {author} {\bibfnamefont {A.~V.}\ \bibnamefont
  {Sergeev}}\ and\ \bibinfo {author} {\bibfnamefont {V.}~\bibnamefont
  {Mitin}},\ }\href {http://link.aps.org/doi/10.1103/PhysRevB.61.6041}
  {\bibfield  {journal} {\bibinfo  {journal} {Phys. Rev. B}\ }\textbf {\bibinfo
  {volume} {61}},\ \bibinfo {pages} {6041} (\bibinfo {year}
  {2000})}\BibitemShut {NoStop}%
\bibitem [{\citenamefont {Klitsner}\ and\ \citenamefont
  {Pohl}(1987)}]{Klitsner:1987}%
  \BibitemOpen
  \bibfield  {author} {\bibinfo {author} {\bibfnamefont {T.}~\bibnamefont
  {Klitsner}}\ and\ \bibinfo {author} {\bibfnamefont {R.~O.}\ \bibnamefont
  {Pohl}},\ }\href@noop {} {\bibfield  {journal} {\bibinfo  {journal} {Phys.
  Rev. B}\ }\textbf {\bibinfo {volume} {36}},\ \bibinfo {pages} {6551}
  (\bibinfo {year} {1987})}\BibitemShut {NoStop}%
\bibitem [{\citenamefont {Karvonen}\ and\ \citenamefont
  {Maasilta}(2007)}]{Karvonen:2007}%
  \BibitemOpen
  \bibfield  {author} {\bibinfo {author} {\bibfnamefont {J.~T.}\ \bibnamefont
  {Karvonen}}\ and\ \bibinfo {author} {\bibfnamefont {I.~J.}\ \bibnamefont
  {Maasilta}},\ }\href {\doibase 10.1103/PhysRevLett.99.145503} {\bibfield
  {journal} {\bibinfo  {journal} {Phys. Rev. Lett.}\ }\textbf {\bibinfo
  {volume} {99}},\ \bibinfo {pages} {145503} (\bibinfo {year}
  {2007})}\BibitemShut {NoStop}%
\bibitem [{\citenamefont {O'Neil}\ \emph {et~al.}(2010)\citenamefont {O'Neil},
  \citenamefont {Schmidt}, \citenamefont {Tomlin},\ and\ \citenamefont
  {Ullom}}]{ONeil:2010}%
  \BibitemOpen
  \bibfield  {author} {\bibinfo {author} {\bibfnamefont {G.~C.}\ \bibnamefont
  {O'Neil}}, \bibinfo {author} {\bibfnamefont {D.~R.}\ \bibnamefont {Schmidt}},
  \bibinfo {author} {\bibfnamefont {N.~A.}\ \bibnamefont {Tomlin}}, \ and\
  \bibinfo {author} {\bibfnamefont {J.~N.}\ \bibnamefont {Ullom}},\ }\href@noop
  {} {\bibfield  {journal} {\bibinfo  {journal} {J. Appl. Phys.}\ }\textbf
  {\bibinfo {volume} {107}},\ \bibinfo {pages} {093903} (\bibinfo {year}
  {2010})}\BibitemShut {NoStop}%
\bibitem [{sup()}]{supplement}%
  \BibitemOpen
  \href@noop {} {}\bibinfo {howpublished}
  {[\url{URL_will_be_inserted_by_publisher}]}\BibitemShut {NoStop}%
\bibitem [{\citenamefont {Tinkham}(2004)}]{Tinkham:2004}%
  \BibitemOpen
  \bibfield  {author} {\bibinfo {author} {\bibfnamefont {M.}~\bibnamefont
  {Tinkham}},\ }\href@noop {} {\emph {\bibinfo {title} {Introduction to
  Superconductivity}}},\ \bibinfo {edition} {2nd}\ ed.\ (\bibinfo  {publisher}
  {Dover},\ \bibinfo {year} {2004})\BibitemShut {NoStop}%
\bibitem [{\citenamefont {Swartz}\ and\ \citenamefont
  {Pohl}(1989)}]{Swartz:1989}%
  \BibitemOpen
  \bibfield  {author} {\bibinfo {author} {\bibfnamefont {E.~T.}\ \bibnamefont
  {Swartz}}\ and\ \bibinfo {author} {\bibfnamefont {R.~O.}\ \bibnamefont
  {Pohl}},\ }\href@noop {} {\bibfield  {journal} {\bibinfo  {journal} {Rev.
  Mod. Phys.}\ }\textbf {\bibinfo {volume} {61}},\ \bibinfo {pages} {605}
  (\bibinfo {year} {1989})}\BibitemShut {NoStop}%
\bibitem [{\citenamefont {Taskinen}\ and\ \citenamefont
  {Maasilta}(2006)}]{Taskinen:2006}%
  \BibitemOpen
  \bibfield  {author} {\bibinfo {author} {\bibfnamefont {L.~J.}\ \bibnamefont
  {Taskinen}}\ and\ \bibinfo {author} {\bibfnamefont {I.~J.}\ \bibnamefont
  {Maasilta}},\ }\href {\doibase 10.1063/1.2357555} {\bibfield  {journal}
  {\bibinfo  {journal} {Appl. Phys. Lett.}\ }\textbf {\bibinfo {volume} {89}},\
  \bibinfo {pages} {143511} (\bibinfo {year} {2006})}\BibitemShut {NoStop}%
\bibitem [{Note1()}]{Note1}%
  \BibitemOpen
  \bibinfo {note} {Static scattering at film boundaries may result from lattice
  distortions due to strain or reconstruction at the film surface and thermal
  expansion mismatch at the film-substrate interface}\BibitemShut {NoStop}%
\bibitem [{\citenamefont {Kaplan}(1979)}]{Kaplan:1979}%
  \BibitemOpen
  \bibfield  {author} {\bibinfo {author} {\bibfnamefont {S.~B.}\ \bibnamefont
  {Kaplan}},\ }\href@noop {} {\bibfield  {journal} {\bibinfo  {journal} {J. Low
  Temp. Phys.}\ }\textbf {\bibinfo {volume} {37}},\ \bibinfo {pages} {343}
  (\bibinfo {year} {1979})}\BibitemShut {NoStop}%
\bibitem [{\citenamefont {Nabity}\ and\ \citenamefont
  {Wybourne}(1991)}]{Nabity:1991}%
  \BibitemOpen
  \bibfield  {author} {\bibinfo {author} {\bibfnamefont {J.~C.}\ \bibnamefont
  {Nabity}}\ and\ \bibinfo {author} {\bibfnamefont {M.~N.}\ \bibnamefont
  {Wybourne}},\ }\href@noop {} {\bibfield  {journal} {\bibinfo  {journal}
  {Phys. Rev. B}\ }\textbf {\bibinfo {volume} {44}},\ \bibinfo {pages} {8990}
  (\bibinfo {year} {1991})}\BibitemShut {NoStop}%
\end{thebibliography}

%

\end{document}